\documentclass[prl,twocolumn,amssymb]{revtex4}
\usepackage{epsfig}
\begin{document}
\title[Ferromagnetism in Graphite]{Ferromagnetism in Oriented Graphite Samples} 
\date{\today}
\author{P. Esquinazi}
\email[E-mail: ]{esquin@physik.uni-leipzig.de}
\author{A. Setzer, R. H\"ohne,  C. Semmelhack}
\affiliation{Department of Superconductivity and Magnetism, Institute for Experimental Physics II, 
University of Leipzig, Linn{\'e}str. 5, D-04103 Leipzig,
Germany}  
\author{Y. Kopelevich} 
\affiliation{Instituto de F{\'i}sica, 
Universidade Estadual de Campinas, 
Unicamp 13083-970, Campinas, S\~{a}o Paulo, Brasil}
 \author{D. Spemann, T. Butz}
\affiliation{Department of Nuclear Solid State Physics,  Institute for Experimental Physics II, University of Leipzig,\\ 
Linn{\'e}strasse 5, D-04103 Leipzig, Germany}
\author{B. Kohlstrunk, M. L\"osche}
\affiliation{Physics of Biomembranes,  Institute for Experimental Physics I, 
University of Leipzig,\\ Linn{\'e}strasse 5, D-04103 Leipzig, Germany}

\begin{abstract} 
We have studied the magnetization of various, well characterized samples of highly oriented pyrolitic graphite (HOPG), 
Kish graphite and natural graphite to investigate the recently reported ferromagnetic-like 
signal and  its possible relation to ferromagnetic impurities. 
The magnetization results obtained for HOPG samples for applied fields parallel to the graphene layers 
- to minimize the diamagnetic background - show no correlation with the magnetic impurity concentration.
Our overall results suggest an intrinsic origin for the ferromagnetism found in graphite.
We discuss possible origins of the ferromagnetic signal.\\

 \end{abstract} 
  \maketitle
\section{Introduction}

We find in the literature reports on ferromagnetism in carbon-based compounds which apparently did
not attract the necessary attention in the community.
For example there are reports on ferromagnetic-like signals in pyrolytic carbon \cite{mizo} and
in HOPG \cite{men}. In the first study  pyrolytic carbon with
higher crystallinity than usual was prepared from adamantane
by chemical vapor deposition \cite{mizo}. The metal impurities were claimed to be $< 25$~ppm and the
ferromagnetic-like loop was reported to persist up to 400 K. In  other work \cite{men} the
main study was the magnetization of Pd nanoparticles encapsulated in a graphite host. 
In the inset of their Fig.~2 the authors show a ferromagnetic-like hysteresis loop for graphite 
without Pd nanoparticles at a temperature of 300~K. We note that the saturation magnetization 
$M_s \sim 10^{-2}$ emu/g is much larger than the one measured in this work 
and by Kopelevich et al. \cite{yakov1}. Because the impurity concentration of that
sample was not reported the origin of this
signal remains, however,  unclear. 
Amorphous-like carbon prepared by direct pyrolysis was found to be a stable strong magnet with a 
saturation magnetization comparable to magnetite \cite{mura1}. The effect of the pyrolysis
temperature on similar materials was also reported \cite{mura2}. Ferromagnetism in  pure C-60 induced by
photo-assisted oxidation was reported to remain up to a critical temperature of  800 K, although the authors
stated that only a small part of the sample contributes to the ferromagnetism \cite{muraka}. More
recently, ferromagnetism in pure rhombohedral polymerized C-60 samples was reported to
remain up to 500 K \cite{maka}.

 The recent report of  ferromagnetic- and superconducting-like \cite{yakov1} magnetization 
hysteresis loops in HOPG  in a broad temperature range  
opens several questions about the origin of the observed anomalies. 
Kopelevich et al. \cite{yakov1} found that the magnetization of HOPG samples shows ferromagnetic-like
hysteresis loops up to 800~K. The details of the hysteresis depend on the sample, sample heat treatment and 
on the direction of the applied field.
There has been a controversy between the opinion of the referees on the possible
influence of impurities on the observations \cite{yakov1}. Therefore,  more systematic work on the observed 
signal was  necessary.

As clearly stated by J.S. Miller \cite{miller} for magnetic polymers, it appears that most of the
claimed  high-temperature ferromagnetism in nominally magnetic-ion free compounds in the past
turned out to be of extrinsic origin. To establish that the low saturation magnetization of a material,
as in the case of the reported HOPG ($M_s < 3$mG $= 3 \times 10^{-3}~$emu/cm$^3$) might be
of intrinsic origin, reproducibility and a quantification of  extrinsic contributions must
be achieved. Therefore and in order to clarify whether the observed ferromagnetism 
found in Ref.~\cite{yakov1} reflects an intrinsic  property of graphite or is due to ferromagnetic 
 impurities, we have  measured  different oriented graphite samples with different 
magnetic impurity concentrations (mainly Fe). 

Both ferromagnetism and superconductivity
have been predicted to arise in graphene layers in regions of certain topological defect structures 
\cite{gon}. On the other hand, ferrimagnetic behavior due to edge states is also expected 
in graphene layers \cite{waka}. 
Taking into account all the experimental and theoretical work cited above
we feel that a systematic study of the ferromagnetic-like hysteresis loops found
in HOPG \cite{yakov1} as a function of magnetic impurities is necessary before speculating on 
its  origin.
Only a careful analysis of the data 
accompanied by a complete impurity concentration study on the same samples
may provide an answer on the origin of the ferromagnetism found in HOPG.

The reported ferromagnetic-like signal  was measured mostly for an applied field in 
direction parallel to the planes, whereas superconducting-like loops in HOPG were mainly observed 
with the magnetic field applied perpendicular to the planes after subtraction of  the diamagnetic 
background \cite{yakov1}. This diamagnetic contribution depends strongly on the angle
between applied field and the graphene layers reaching a minimum for the parallel orientation.  
Therefore, and in order to minimize any possible error due to the subtraction of 
the diamagnetic contribution, all the results
reported in the current study were obtained with the field parallel to the graphene layers. With respect to the
superconducting-like behavior observed in graphite \cite{yakov1} we would like to 
note that clear evidence for superconductivity in simple mixtures of graphite 
powder with sulphur  has been reported recently \cite{yakov2}.

 The paper is organized as follows. In the next Section the samples characteristics and experimental 
details are given. In Section 3 we describe  the results and in Section 4 we discuss the observed 
behavior and its possible origin. A short summary is given in Section 5.  

\section{ Sample  Characteristics and Experimental Details}

 Table I shows the magnetization  
characteristics, the  impurity concentrations,  the masses as well as the full width at half maximum (FWHM) of 
x-ray rocking curves in $\theta$ ($\Delta \theta$) and in $2\theta$ ($\Delta (2\theta$)) of the measured
 graphite samples. 
Samples HOPG-2 and -3 were prepared at the Research Institute ``Graphite" (Moscow)
as described in Ref.~\cite{brandt}. Two commercial  HOPG samples were from 
Union Carbide (samples UC3 and UC4,  cut from the same batch) and  Advanced
Ceramics (samples AC1, batch 10158, and AC2, batch 25672). Furthermore, we have characterized
a Kish and a natural graphite crystal, this last obtained from Ticonderoga (USA) \cite{munich}.

 The magnetization measurements were done with a Superconducting Quantum  Interference 
Device (SQUID) from Quantum Design with a 7~T superconducting solenoid. 
In general we have selected three cm scan length for the
magnetization measurements. The SQUID response enables measurements of samples with a 
minimum magnetic moment $m  \sim 3 \times 10^{-10}~$Am$^2~  ( 3 \times 10^{-7}$ emu).

Because the magnetization signals of the samples are rather small we have checked the 
background signal from the sample holders in the SQUID. This 
contribution turned out to be
non-negligible,
and special care has been applied in the selection of the holder
as well as  with the sample handling procedure in order to prevent contamination
of the holders and/or samples. For measurements below 350 K we 
have used a plastic tube with a small window and an adhesive tape (Tesa AG, Hamburg). 
The background signal of this
sample holder with tape but without sample is shown in Fig. 1. 
Small but measurable hysteresis is observed which lies
close to the resolution limit of our SQUID magnetometer.
 For measurements above 350 K the samples were 
fixed in a quartz tube between two quartz spacers and enclosed under a pressure
of $2 \times 10^{-6}$ torr. The background signal in this case was 
 negligible in comparison with the sample response.

In general, the ferromagnetic hysteresis is superimposed to a linear in field diamagnetic
background. As noted above, this diamagnetic
signal depends on the misalignment of the sample with respect to the field as
well as on the intrinsic misalignment of the graphene planes in the sample.
The diamagnetic contribution for each sample has been determined in-situ and 
in a straightforward way: Using a linear
regression through the data points obtained for magnetic fields up to 5~T we obtained
the diamagnetic slope $\chi_{||} = M(H) / H$. We note that the magnetization
due to the superimposed ferromagnetic signal saturates at fields larger than 2~kOe and
therefore it does not influence the linear field dependence of the diamagnetic background. 
The obtained susceptibility  ranges between $0 \ge \chi_{||}(10 K) \ge -5.3 \times 10^{-6}~$emu/gOe for
the studied samples. The ferromagnetic loop is then obtained from the measured
magnetization $M(H)$ subtracting the diamagnetic background as $M(H) - \chi_{||} H$.

The crystallinity of the graphite samples was investigated by high resolution x-ray diffraction
with a Philips X'pert diffractometer.
$\theta$ and $2\theta$ scans around the symmetrical (002) reflection were carried out
using the Cu-K$\alpha$1 radiation. Furthermore, for three samples we have determined the average in-plane
correlation length $D$ from the $2\theta$ scans around the symmetrical (110) reflection. As
expected, $D$ correlates with the 
FWHM
 obtained from the $\theta$ (002)-scans. Table I shows the FWHM of the 110 reflection as well
as $D$ for the samples AC2, HOPG-2 and HOPG-3.

The measurements of the impurity concentration were carried out with Particle
Induced X-ray Emission (PIXE) using a 2 MeV proton beam.
For the samples studied here the typical minimum detection limit (MDL) of 
this method 
was $\le 2 \mu$g/g (for some elements, much smaller). With exception of the natural
graphite sample, the impurity concentration was distributed homogeneously
in the samples. The concentration of all other magnetic elements not listed in Table~I was
less than 1 $\mu$g/g.

Scanning tunelling microscopy (STM) was performed with a DI Nanoscope E in 
conjunction with a Small Sample scan head (``A" scanner, Digital Instruments, 
Santa Barbara, CA). 
Pt/Ir (80:20) tunneling tips (wire diameter: 0.25~mm) were from Plano (Wetzlar, Germany). 
Images were obtained under ambient conditions with a tip bias of typically 
$+20$~mV in the constant current-mode 
(scan rate $f = 3 \ldots 6~$Hz; current $I_0 \approx 2~$nA) or in the constant height-mode 
(scan rate $f = 20 \ldots 60~$Hz). While in both modes, a clear-cut 
discrimination of topology (height) differences from density-of-state (current) 
differences cannot be afforded, the first scanning mode emphasizes the 
topological structure whereas the latter accentuates local variations of the 
density of states.

\section{Results}

Figures \ref{ac} to \ref{hopg2} show examples of the measured magnetization loops  
  obtained in some of the samples at 300 K and
 10 K.  All HOPG samples
show ferromagnetic-like hysteresis of the same order of magnitude.
Figure \ref{ac} shows the hysteresis loops obtained for the samples from
Advanced Ceramics at two different temperatures. Whereas the loop shown in 
(b) is obtained after subtraction of a  diamagnetic background signal $\chi_{||} = - 1.32 \times 10^{-6}~$emu/g Oe,
the loop  in (a) was obtained directly from the measurements without subtraction of
any diamagnetic background. This result and the range of values obtained for $\chi_{||}$ suggest
that the diamagnetic signal measured at high fields applied nominally parallel to the planes of HOPG samples 
is strongly influenced by the
misalignment of the sample.

Figure \ref{uc} shows the loops obtained for two Union Carbide samples of different masses cut from the
same batch. These results provide a measure of  the typical scatter of the main values of the hysteresis
loops in similar samples.

In order to check if there is any correlation of the hysteresis values with the measured 
Fe-concentration we plot in Figs.~\ref{rem}, \ref{sat} and \ref{hc} the remanent magnetization $M_{r}$,
the magnetization at 2kOe $M(2$kOe) and the coercive field $H_c$, respectively, as
 a function of impurity content. From  Fig.~\ref{rem} 
one would tend to infer
a weak increase of $M_{r}$ with Fe concentration. However, 
the remanent magnetization is not the characteristic quantity that one expects to change systematically with
Fe concentration since it depends primarily on the pinning of magnetic domains. Rather, 
a correlation of the saturation magnetization with the impurity content would be significant. 
Figure~\ref{sat} shows that such a correlation is not observed within experimental error. 
A similar result is obtained for the coercive field, see Fig.~\ref{hc}.

The FWHM obtained in $\theta$ and $2\theta$ scans provides
a measure of the variation in the orientation of a particular set of lattice planes 
(mosaic spread) and in the spacing  of the graphene planes (e.g., inferred by dislocations), 
respectively.
Figure \ref{fwhm} shows the magnetization at saturation as a function of $\Delta \theta$ (a)
and $\Delta (2 \theta$) (b). We find no systematic dependence of $M(2$kOe) on $\Delta \theta$.
A clear  increase of $M(2$kOe) with $\Delta (2 \theta)$ can be recognized in Fig.~\ref{fwhm}(b).

It has been reported \cite{yakov1} that annealing of HOPG
samples leads to the disappearance of  the superconducting-like hysteresis loop and turns
 it into a ferromagnetic-like
loop. Also, short time annealing ($\sim 2~$h) increases the ferromagnetic-like signal \cite{yakov1}. 
In the current work   we have measured the hysteresis as a function of the annealing
time and temperature. 
An increase of the ferromagnetic signal at the beginning of the annealing procedure was also observed in this work.
Longer time annealing reduces however the ferromagnetic signal and reaches similar values 
as before the annealing. 
In Fig.~\ref{hopg2}(b) we show the data obtained
at 300~K after annealing a different piece of  the HOPG-2 sample for
16 h at 700~K ($\circ$) and additionally after 19 h at 
800 K $(\blacktriangle)$. The annealing procedure was 
done in situ in the SQUID under vacuum. It is observed that long time annealing has a weak
influence, if any,  on the hysteresis width and the saturation magnetization.

\section{Discussion}

The origin of  the ferromagnetic-like signal in HOPG samples is controversial.
Therefore, we will discuss below different possible origins in view of 
the existing evidence.

{\it (a)  Magnetic impurities}.
The possibility of a significant contribution of magnetic impurities to the measured  magnetic properties is taken serious and 
carefully analyzed before we discuss possible intrinsic origins of the hysteresis loops measured in graphite. 
In spite of the fact that no correlation between the observed magnetization
 and the Fe-concentration exists, one
might still believe that those impurities are sufficient to produce the measured signals. 
A typical argument used to attribute a ferromagnetic signal to impurities would  be as follows: 
Let us assume  that all Fe impurities  would form iron or magnetite (Fe$_3$O$_4$) clusters  and that all 
clusters are large enough to behave ferro- or ferrimagnetically. 
An impurity content  of 1~$\mu$g Fe per gram graphite would then 
contribute $\simeq 2.2 \times 10^{-4}~$emu/g  to the magnetization 
in the case of Fe clusters or  $\simeq 1.4 \times 10^{-4}~$emu/g
in the case of Fe$_3$O$_4$ clusters assuming that these clusters behave like bulk materials.
Considering a linear increase of the magnetic moment with the Fe concentration, 
for most of the samples the possible contribution of the Fe impurities 
exceeds clearly  the measured magnetization values at saturation
or are on the same order of magnitude, see Fig.~\ref{sat}.
In the case of the Kish sample, AC2  and for HOPG-3 the measured values are larger than the
maximum possible contribution from Fe clusters. Thus, for these three samples
the values of magnetization cannot be explained by Fe impurities. 

Moreover, the assumption that such a low Fe-concentration behaves 
ferromagnetically and produces a hysteresis loop with finite remanent moment and 
coercive fields at room temperature or even higher \cite{yakov1} is
 rather unlikely taking into account that the Fe concentration measured in the HOPG
samples was found to be homogeneously distributed. To quantify the magnetization  behavior of
Fe impurities in graphite we have measured a natural graphite bulk sample that has an inhomogeneous
Fe distribution through the sample between 13 $\mu$g/g and 3,790 $\mu$g/g;  most of the sample 
shows a concentration larger than 500~$\mu$g/g Fe. All other 
magnetic impurities are below 10 $\mu$g/g and may therefore be neglected.
 Figure \ref{nat}(a) shows the hysteresis loop of  the natural graphite crystal for fields parallel to its 
main plane after subtracting the paramagnetic (10 K) or diamagnetic (300 K) contributions
(proportional to the applied field) shown in the inset. 
The measured hysteresis is very small with a remanent magnetization 
$M_{r}< 2 \times 10^{-5}$~emu/g. The magnetization at 2kOe is $< 4 \times 10^{-3}$~emu/g. Naively, one
would argue that such a magnetization  can be produced  by $\sim 18$~$\mu$g/g ferromagnetic Fe. 
Such a comparison is, however, misleading. 
If the magnetic signal is due to Fe impurities, the nearly negligible hysteresis width
and the lack of saturation at 2~kOe indicate that  Fe impurities do not behave  ferromagnetically but  rather 
superparamagnetically. To better understand this point 
we measured the magnetization up to 5~T and we obtained a ferromagnetic-like curve of negligible width, see Fig.~\ref{nat}(b),
 after subtracting the paramagnetic
background contribution.
We estimate that the saturation magnetization observed at 5~T is 
produced by 840~$\mu$g/g  Fe, a concentration 
comparable to the average 
impurity content of the sample. 

A typical response of superparamagnetic  Fe in graphite is also recognized in the strong temperature
dependence of the magnetization at 0.2~T below 150~K, see inset in Fig.~\ref{nat}(b). This
behavior is in clear contrast to that obtained in the HOPG samples. As an example we show
the temperature dependence of the magnetization measured at 2kOe
in  the HOPG-2 sample (after subtracting the diamagnetic contribution) 
 in Fig.~\ref{magt}. It is recognized that
 the magnetization is practically temperature independent, in clear
contrast to the expected behavior if the 8 $\mu$g/g of Fe would behave superparamagnetically.
This result agrees with that reported in Ref.~\cite{yakov1} that the saturation magnetization 
 in annealed HOPG samples remains practically
temperature independent up to 800~K. 
In line with that, the measured hysteresis loops for the HOPG-2 sample at 800~K 
(which is the highest possible in our equipment) indicate that the
Curie temperature is larger than that. From a linear extrapolation of the remanent 
magnetization (Fig.~\ref{magt}) to zero we estimate
that the Curie temperature is of  the order of 900~K. 
 The small change observed in the hysteresis upon annealing the sample at 700 K or 
 800 K in vacuum,
shown in  Fig.~\ref{hopg2}(b), 
appears to be difficult to explain by arguing in terms of a magnetic contribution of ferrocene.  
From all the evidence presented here we believe that magnetic impurities can be 
excluded as the origin of the weak ferromagnetism observed
in HOPG . 

{\it (b) Topological defects: Grain boundaries and edge states}.
The topological and electronic structure of atomically flat graphite surfaces 
has been extensively studied by scanning tunneling microscopy (STM) 
\cite{Binnig,Quate,Soler,Fuchs}. 
More recently, the electronic properties of carbon nanotubes, and particularly 
edge states of nanometer-sized graphite ribbons, have attracted much interest 
\cite{Nakada,giu}. It is expected that the electronic structure of the graphite 
$\pi$ system may differ considerably on carbons located within a graphite sheet 
and those at its edge \cite{fuji2,fuji1}. Moreover, such localized edge states 
have been shown to depend sensitively on the geometry of the edge \cite{Nakada}: 
While ``zigzag" edges perpendicular to the $[1,0,\bar{1},0]$ lattice direction show 
such states, ``armchair" edges (perpendicular to $[1,1,\bar{2},0]$) do not. 
Thus, ``zigzag" edge states may lead to an increase of the density of states at 
the Fermi level. 
This has been experimentally confirmed in STM work on HOPG samples \cite{giu}.
If such edge states occur at high density, a ferrimagnetic spin polarization 
may result. Experimental findings on graphitized nanodiamond \cite{and} 
and activated carbon fibers \cite{nak} support this expectation. 
Similarly, a strong enhancement of the Curie-like paramagnetic 
contribution was predicted at low temperatures, see Ref.~\cite{waka} and references 
therein, and this has been experimentally confirmed \cite{shi}.
Recently, the mechanisms of magnetism in stacked nanographite were theoretically 
studied \cite{hari}. This work obtained an antiferromagnetic solution for
A-B-type stacking only. 

One may doubt that the edge states could be the origin for the 
 ferromagnetism found in our macroscopic graphite samples, since
in nanographites these states  provide an enhanced paramagnetism
 and their size $\sim 2~$nm is smaller than the in-plane correlation length of our graphite
samples (the influence of  the edge states
to the electronic properties of a graphene layer should decrease with $D$).
However, small-size effects and the thermal energy could also preclude the formation 
of a stable ferromagnetic
ordering in nanographite. It is also unclear whether correlation effects between
edge states  in macroscopic graphite samples with a
in-plane correlation length not more than five times the size of the reported nanographite    
samples (see Table I)  would not be enough to stabilize a ferromagnetic 
order. Theoretical studies suggest that the presence of strong topological disorder
will stabilize ferromagnetism and frustate antiferromagnetic order \cite{gon}. No 
prediction was made, however, on the  minimum density of the disorder necessary for this situation . 
Therefore and because the theoretical results are not yet conclusive we should not
neglect a possible contribution of edge states to stabilize the FM in bulk graphite.

In order to assess the role of edge states and their potential contribution 
to the magnetic properties of the graphite samples studied in this work, a 
systematic STM investigation of these samples would be desireable. Unfortunately, 
such a study is beyond the scope of our current experimental capabilities. 
In order to obtain at least a feeling whether or not edge states that might exist 
at grain boundaries within the graphite samples contribute to their observed 
anomalous magnetic properties we have conducted a limited study of some of the 
samples in air. We note that we are currently not capable of performing the 
spectroscopic studies \cite{Reihl} that were required to assess quantitatively 
putative differences in the density of states at the Fermi level as a function 
of the atom positions with respect to grain boundaries.

Figure~\ref{stm} gives exemplary STM results obtained at a freshly cleaved 
surface of the HOPG-2 sample in air. The images have been flattened, but no 
filtering was applied to the raw data. Panel (a) shows a $125 \times 125~$nm$^2$ 
surface topology overview observed in constant-current mode. 
Two types of linear defects are observed within the planar surface: 
A step line, oriented roughly along the scan direction (horizontal direction 
within the image) and an apparent elevation line that runs along the step defect 
at an oblique angle. Both line defects are much longer than the width of the image 
and are characterized in the section profile shown on the right. 
In other micrographs than the one shown in Fig.~\ref{stm} we do not find 
indications that the two type of line defects are coupled to each other.
By inspection of a large number of similar micrographs, we conclude that
such defects are characteristic of the sample surface. On the surface of the 
HOPG-2 sample, they have a typical distance of a few 100~nm. 
Inspection of the unit cell orientations on both sides of the two types of 
defects at higher resolution ({\em vide infra}) reveals that both are 
associated with grain boundaries.

The section profile indicates that the step height of the first line defect 
is $\approx 0.65~$nm, which corresponds to the length of one unit cell 
along the crystallographic {\bf c} axis, {\em i.e.} the height of two graphite 
sheets. In fact, an additional feature along the defect line at mid-level of 
the step is recognized. Since two stacked $\pi$ rings in the unit cell are 
displaced sideways with respect to each other in the graphite crystal, this half-step 
may correspond to an individual graphite layer that  protrudes laterally 
underneath the top layer. At a distance of a few nm from the step line, the 
profile is essentially flat. 
This is expected as the two planes correspond presumably to equivalent crystal 
surfaces that differ by one unit cell in height.
However, as the trace on the higher level approaches the step, a pronounced 
steepening of the slope is observed. Since the scan direction was chosen to be 
parallel to the step line, it cannot be argued that this feature is due to tip overshoot. 
Rather, because the experiment is inherently incapable of discriminating between 
topological features and changes of the density of states, the observed 
steepening of the contour line may more likely be associated with an increase of 
the local density of states on the carbon atoms as the topology approaches 
the grain boundary.

The apparent topology of the second line defect is distinctly different. 
The section profile registers a sharp, symmetric peak that has an apparent 
height, $\Delta z \simeq 0.25~$nm, and an apparent width of $\approx 2~$nm. 
At higher resolution (Fig.~\ref{stm}(b), showing a constant-height scan of 
a similar defect on the same sample surface) the defect structure is disclosed 
in more detail. As recognized in the overview (inset) the defect line appears as 
three collinear stripes. These are no tip 
artifacts since they do not depend on the relative orientation between the 
defect line and the scan direction. The high resolution micrograph reveals 
that the corrugation is associated with individual surface atoms 
and that the lattice orientation differs on both sides of the central spine.
The apparent height of the features with the highest elevations -- truncated 
in the grey level scale which was chosen to better visualize the 
atomic lattice on the surrounding crystal faces -- is about a factor of two 
larger than an atomic diameter. 
It is thus likely that an increase in the density of states on atoms near the 
grain boundary contribute to the appearance of these corrugations.
Without a clean discrimination of topological and 
density-of-state features, however, it cannot in this case be decided with 
confidence whether or not the observed defects correspond simply to lines of 
atoms that decorate the grain boundary.

A high-resolution characterization of step-type line defects is presented in the 
lower half of Fig.~\ref{stm}. Panels (c) and (d) show the vicinity of the 
defect shown in Fig.~\ref{stm}~(a) at high resolution in 
constant-current and constant-height mode, respectively. The scan direction 
has been rotated by $\approx 90^o$ with respect to that in the overview scan. 
Similar to Fig.~\ref{stm} (b), the images show the surface at atomic 
resolution. The fundamental corrugation is due to a 
carbon site asymmetry \cite{Binnig,Quate} giving rise to a hexagonal lattice.
Again, the constant-current micrograph -- Fig.~\ref{stm}(c) -- indicates a step 
height of $\Delta z \approx 0.6~$nm (section profile on the right).
Both micrographs reveal clearly a misorientation of the lattices on both sides 
of the defect line of $42^o \pm 5^o$. On the higher level -- to the left -- 
the main lattice direction meets  the grain boundary at $65^o \pm 5^o$, 
characteristic of a ``zigzag" edge. By contrast, one would expect this angle to 
be $90^o$ for an ``armchair" edge. In distinction from an earlier report 
\cite{giu}, where ``zigzag" edges were observed to extend only for short 
lengths, the defects observed here may extend for $> 100~$nm, as 
demonstrated in Fig.~\ref{stm}~(a) where the step defect runs along the same 
general direction over the entire, 125~nm, width of the micrograph.

In the close vicinity of the grain boundary itself, the constant-height image, 
Fig.~\ref{stm}~(d), appears distinctly different from the constant-current image. 
It is characterized by a vast overshooting of the section profile at the 
locations of the edge carbon atoms, {\em c.f.} pseudo-3D representation on the 
right of Fig.~\ref{stm}(d). While we reiterate that a clean distinction of 
topological and density-of-state features is not afforded in our measurements, 
we may note that the observed effect cannot easily be accounted for by the 
sample topology or elasticity effects \cite{Soler}. For example, we consider it 
unrealistic that a decoration of the grain boundary with atoms (of an ordinary 
-- low -- density of states at the Fermi level) should be responsible for the 
observed giant corrugations. Rather, a large increase of 
the local density of states at the atom positions near the edge must be invoked 
to explain the observations. This is also hinted by the deviation of the 
section profile in the overview scan, Fig.~\ref{stm}~(a), from a planar topology near the step. 
The conjectured enhancement of the density of states 
on edge atoms may indicate the existence of a localized state as demonstrated in 
simulations \cite{koba}. Therefore, one may argue that topological defects such as the 
``zigzag" edge characterized in Fig.~\ref{stm} affect the electronic properties 
of the bulk graphite crystal and might also be responsible, at least partially, 
for the measured ferromagnetic  behavior. Clearly, a 
more definite conclusion on  the importance of such localized states requires 
much more systematic STM work. In future, we are also planning to vary 
systematically the number of topological defects and thus quantify their 
influence on the magnetization. Preliminary results indicate indeed an increase 
of the ferromagnetic signal upon milling of a bulk sample. 

Finally, we note that recent theoretical work points out the effects of the 
electron-electron interactions in a graphene layer suggesting the existence of 
both ferromagnetism and an anisotropic $p-$wave superconducting state \cite{gon}. 
This work also suggests that topological disorder in the graphene planes may 
enhance the electronic density of states and induce instabilities in the 
electronic system, thereby giving rise to the behavior reported here and in 
earlier work \cite{yakov1}.

{\it (c) Itinerant ferromagnetism}.
 Since the carrier density in graphite is low, the magnetic properties 
may be connected to itinerant ferromagnetism of dilute two-dimensional electron gas systems
when the electron-electron interaction is large. 
Indirect evidence  for the importance of the electron-electron interaction in graphite
 has been recently obtained by magnetoresistance measurements that 
showed a field induced metal-insulator like 
transition with 
a scaling quantitatively similar to those found in MOSFET's and amorphous thin films \cite{yakov3}.  

In general, the strength of  the electron-electron interaction can be estimated through 
the Coulomb coupling constant
\begin{equation} 
r_s = \frac{1}{(\pi n_{2D})^{1/2} a_B^\ast}\,,  
\label{rs} 
\end{equation} 
where the 2D carrier density is given by $n_{2D} = n_{3D}d$ with the interplane 
distance $d=0.335~$nm, the effective Bohr 
radius $a_B^\ast = \epsilon \hbar^2/e^2 m^\ast$, and  the dielectric
 constant for graphite $\epsilon = 2.8$. The 3D carrier density of the majority band is 
$n_{3D} \simeq 2 \times 10^{18}~$cm$^{-3}$ with an effective mass 
$m^\ast \sim 0.05~m_0$ where $m_0$ is the 
free electron mass. For the minority carriers, $n_{3D} \simeq 6 \times 10^{16}~$cm$^{-3}$ with an 
effective mass $m^\ast \sim 0.004~m_0$ \cite{dre2}. Substituting these values in eq.~(\ref{rs}) we 
obtain $r_s \sim 5$ and $\sim 10$ 
for the majority and minority carriers, respectively.  
A value of $r_s$ larger than 1 implies that the potential energy per 
electron is larger than the Fermi energy, hence 
effects due to the electron-electron interaction must  be 
taken into account.

If we assume that Dirac fermions are responsible for the magnetic 
field driven metal-insulator-like transition in
graphite (for fields parallel to the $c-$axis), the strength of the Coulomb 
coupling is characterized by the massless and dimensionless 
parameter \cite{khv,voz}
\begin{equation}
g = \frac{e ^2}{\epsilon_0 v_F \hbar}\,,
\label{g}
\end{equation}
where $e$ is the electronic charge, $\epsilon_0$ the dielectric constant and $v_F \sim 2 \times 10^6$~m/s the 
Fermi velocity for graphite. As noted recently \cite{khv},
for graphite, $g > 10$ holds which suggests that the strong Coulomb interaction can open an
excitonic gap in the spectrum of Dirac fermions. Interestingly, the occurrence of
the gap is accompanied by the appearance of a small magnetic moment as the result of  
band anisotropy \cite{khv}. In agreement
with this theory, conduction electron spin resonance (CESR) experiments performed in
HOPG and Kish graphite samples reveal a ferromagnetic-like internal field \cite{ser}.
It is expected that this doped excitonic FM would disappear above a certain level of
doping of the order of the chemical potential \cite{khv}. We note  that
within this picture the excitonic gap and the nonzero spin polarization is driven
by the Coulomb interaction even at zero magnetic field. On the other hand, a field 
applied normal to the graphene layers can induce the formation of a gap 
even when the Coulomb interaction is weak \cite{khv}, a model used to interpret the recently 
found metal-isolator transition in HOPG \cite{yakov3}.

It is also interesting to compare the magnetism found in graphite with that in the 
high-temperature itinerant Ca$_{1-x}$La$_x$B$_6$ ferromagnet \cite{you} 
where either quantum localization effects \cite{cep} or an excitonic mechanism 
\cite{gor} are invoked to explain the observed 
weak ferromagnetism with a Curie temperature of 600~K . 
In graphite the low-temperature ferromagnetic magnetization at saturation
is $\sim 1 \times 10^{-3}~$emu/g. 
This value can be  translated in terms of the magnetic moment $\sim 0.1 \mu_B$ per majority carrier
similar to $0.07 \mu_B$ found in Ca$_{1-x}$La$_x$B$_6$.  According to
Ref.~\cite{khv} the ferromagnetism in graphite and hexaborides can have the same origin.
Direct experimental evidence supporting an excitonic mechanism for the FM found in
HOPG is, however, still lacking.
 
\begin{center} 
{\bf 5. Summary} 
\end{center}
 We have studied the magnetization of several HOPG, Kish graphite and natural graphite samples with
different content of magnetic impurities. Our results rule out that ferromagnetic impurities can be 
 responsible  for the observed magnetic effects in HOPG and Kish graphite. 
The magnetization at saturation remains practically
temperature-independent up to 500~K. 
This behavior is in clear contrast to that found in natural graphite samples with at least two
orders of magnitude larger concentration of Fe-impurities. The results of natural graphite
reveal the superparamagnetic behavior of Fe impurities in the carbon matrix.
Long-time annealing at 700~K and 800~K in vacuum of the HOPG samples affects the hysteresis loops 
only weakly. 

The origin of the FM in HOPG is not yet clear. We have discussed two possible origins in this 
work. Topological defects can contribute to the peculiarities of the graphite
electronic properties and  may give rise to ferromagnetic correlations. Also, the strong Coulomb 
interaction between electrons in graphite (due to the small electronic density) should play an important role in all 
the observed magnetic properties of graphite. However, details of the long-range
itinerant ferromagnetic order remain unclear. 

The discovery of ferromagnetism
in oxidized C$_{60}$ \cite{muraka} as well as in polymerized rhombohedral C$_{60}$ \cite{maka} 
indicates that carbon-based materials can be ferromagnetic without metallic components at high temperatures. 
Therefore, it should be no surprise if
graphite itself shows weak ferromagnetic signals.
Future work should try to (a) enhance the ferromagnetic signals by an appropriate mixing of graphite
with other elements (an enhancement of the superconducting 
signal has been recently achieved mixing graphite with sulfur \cite{yakov2,yang}), 
(b) increase  the defect concentration as well as (c) prove
experimentally the existence of an excitonic gap. 

\begin{acknowledgements}
  
We thank J. G. Rodrigo for  the comments on our  STM studies. 
This work is supported by the Deutsche Forschungsgemeinschaft within  DFG Es 86/7-1. Y.K. was 
also supported by CNPq, FAPESP and CAPES. We acknowledge the support of the DAAD.  
M.L. is supported by the Fonds der Chemischen Industrie, Frankfurt.
\end{acknowledgements}


\begin{widetext}

\newpage

\begin{table}
\caption{Characteristics of the investigated samples. The samples AC1, AC2, UC3, UC4, HOPG-2 and HOPG-3 are
highly oriented pyrolytic graphite (see text), NG is a natural graphite crystal from Ticonderoga (USA) \protect\cite{munich}.
${(a)}$ All magnetization values are in units of $10^{-4}$ emu/g. The magnetization values in the table were 
obtained after subtraction of  a  linear background contribution.
 (b) The Fe concentration in this sample is not homogeneous. In four different position of the sample 
the Fe concentration was found to be between 0.10\% and 0.38\% (wt), in two other positions the values were between 
13 $\mu$g/g and 165 $\mu$g/g. (c) All impurity concentration values are in $\mu$g/g. For all samples 
the MDL $\le 2~\mu$g/g. n.d.: not determined.}
\begin{tabular}{|l|c|c|c|c|c|c|c} 
Sample: 		& AC1		 & AC2 	& UC3/UC4 	& HOPG-2 	& HOPG-3 	& Kish 		& NG 	 \\ \hline
mass (mg): 	& 26.5  		 & 18.9	& 15.5/9.9	& 23.7 		& 55.1 		& 2.9 		&  21.7	  \\ \hline
FWHM ($\theta$,002)& 0.78$^o$ 	& $0.401^o$  	&0.237$^o$	& 1.33$^o$	& $0.60^o \pm 0.05^o$ &$\sim 1.6^o$ & 5.05$^o$ \\ \hline
FWHM ($2\theta$,002)&0.124$^o$  	& $0.116^o$	& 0.119$^o$	& 0.118$^o$	& $0.22^o$	&0.13$^o$	& -	 \\ \hline
FWHM ($2\theta$,110)& n.d. & $0.966^o$ & n.d. & $2.08^o$ & $1.174^o$ & n.d. & n.d. \\ \hline
$D$ (nm) & n.d. & $11 $ & n.d. & $5$ & $9 $ & n.d. & n.d. \\ \hline
$M(2$kOe,10K)$^{(a)}$ & n.d. & $2.7 \pm 0.2 $ & $5.6 \pm 0.4/12\pm 2$  & $9.7 \pm 1$  & 25 & $10 \pm 2$ & $7.8 \pm 0.2$   \\ \hline
$M(2$kOe,300K) & $13.9$& $3.3 \pm 0.4$  & $8.3 \pm 0.3 / 11 \pm 2$ &$9.3 \pm 1$  & 25 & $6 \pm 2$ & -  \\ \hline
$M_{r}(0, 300$K) & 3.58 & $0.5\pm 0.1$ & 2.2/2.8 & $1.2 \pm 0.2$  & 1.77 & $0.7 \pm 0.2$ & -   \\ \hline  
$H_c(10$K) (Oe) & n.d. & $170 \pm 30$  & $100 \pm 10 / 94 \pm 5$ & $105 \pm 10$ & $85 \pm 2$& $170 \pm 20$ & $< 50$   \\ \hline
$H_c(300$K)(Oe) & 98  &$70 \pm 10$  & $70 \pm 5 / 65 \pm 10$ & $58 \pm 4$ & $60 \pm 2$ & $80 \pm 2$ & -  \\ \hline
Fe$^{(c)}$  & $13 \pm 1$ & $ < 0.3 $  &  $16 \pm 2$ & $8 \pm 2$ & $ 7.8 \pm 2$ & $0.7 \pm 0.3$ & ${(b)}$  \\ \hline
Ni  &  $1.3 \pm 0.5$  & $< 0.4$  & $2.1 \pm 0.7$ & $0.8 \pm 0.5$ & $< 0.2$ & $< 0.4$ & $ 3 \pm 2$   \\ \hline
Mn & $< 0.3$ & $< 0.3$  & $< 0.2$ & $< 0.2$ &$<0.2$ & $< 0.3$ & $10 \pm 5$  \\ \hline
Cu & $16 \pm 2$ & $< 0.5$ & $73 \pm 5$ & $1.2 \pm 0.5$ & $1.3 \pm 0.5$ & $< 0.5$ & $3 \pm 2$  \\ \hline
Ca & $< 0.7$ & $< 0.7$  & $56 \pm 6$ & $2.6 \pm 0.8$ & $9.7 \pm 2.5$ & $< 0.7$ & $100 \pm 30$  \\ \hline
Ti & $9 \pm 2$  & $7 \pm 1$ & $3.3\pm 1$ & $0.8 \pm 0.5$ & $1.0 \pm 0.7$ & $< 0.5$ & $< 1$ \\ \hline
\end{tabular}

\label{samples}
\end{table}

\newpage
   
\begin{figure} 
\centerline{\mbox{\epsfysize=15cm \epsffile{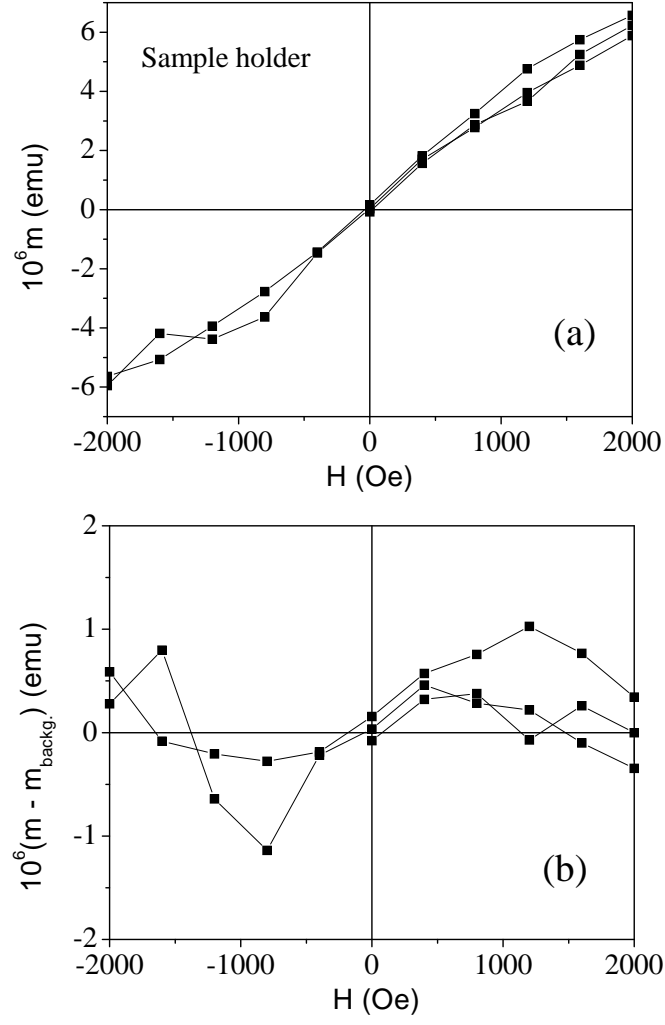}}} 
\caption{(a) Magnetic moment of the sample holder (plastic tube with a small window and an
adhesive tape) as a function of applied field at $T = 10~$K. (b) The results from (a) 
after subtraction of the linear 
paramagnetic background. 1 emu/g = 1 A m$^2$ / kg, 1 Oe = $10^3 / 4 \pi$ A/m.} 
\label{1} 
\end{figure} 

\begin{figure} 
\centerline{\mbox{\epsfysize=15cm \epsffile{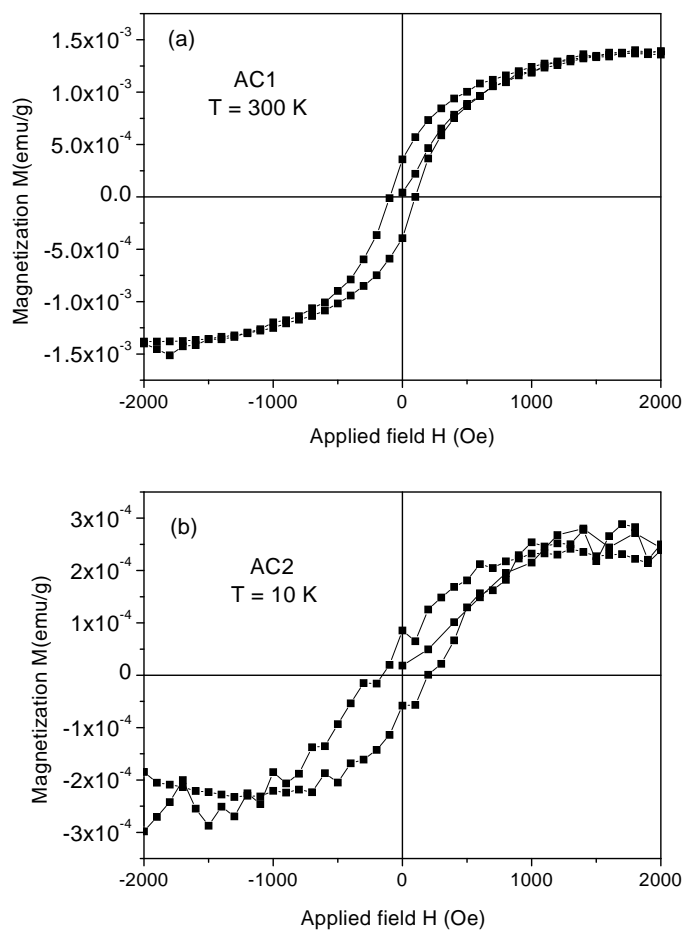}}} 
\caption{(a) Magnetization as a function of applied field at $T = 300~$K for sample AC1. 
No background subtraction has been performed. (b) The same for sample AC2 at 10~K after
subtraction of a diamagnetic background, $\chi = - 1.32 \times 10^{-6}$ emu/g Oe. The sample
holder background (see Fig.~\protect\ref{1}) was not subtracted from the data.} 
\label{ac} 
\end{figure} 

\begin{figure} 
\centerline{\mbox{\epsfysize=15cm \epsffile{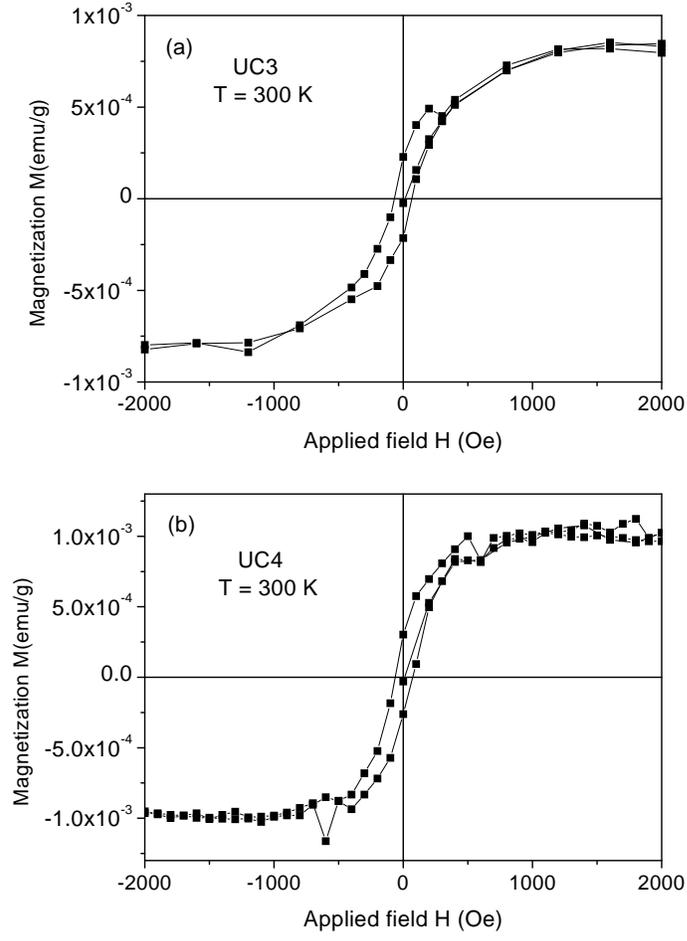}}} 
\caption{(a) Magnetization as a function of applied field at $T = 300~$K for sample UC3 (Union Carbide). 
A diamagnetic background $\chi = - 2.58 \times 10^{-6}$ emu/g Oe has been subtracted. (b) 
The same for sample UC4 at 300~K after
subtraction of a diamagnetic background $\chi = - 1.74 \times 10^{-6}$ emu/g Oe.} 
\label{uc} 
\end{figure} 

\begin{figure} 
\centerline{\mbox{\epsfysize=15cm \epsffile{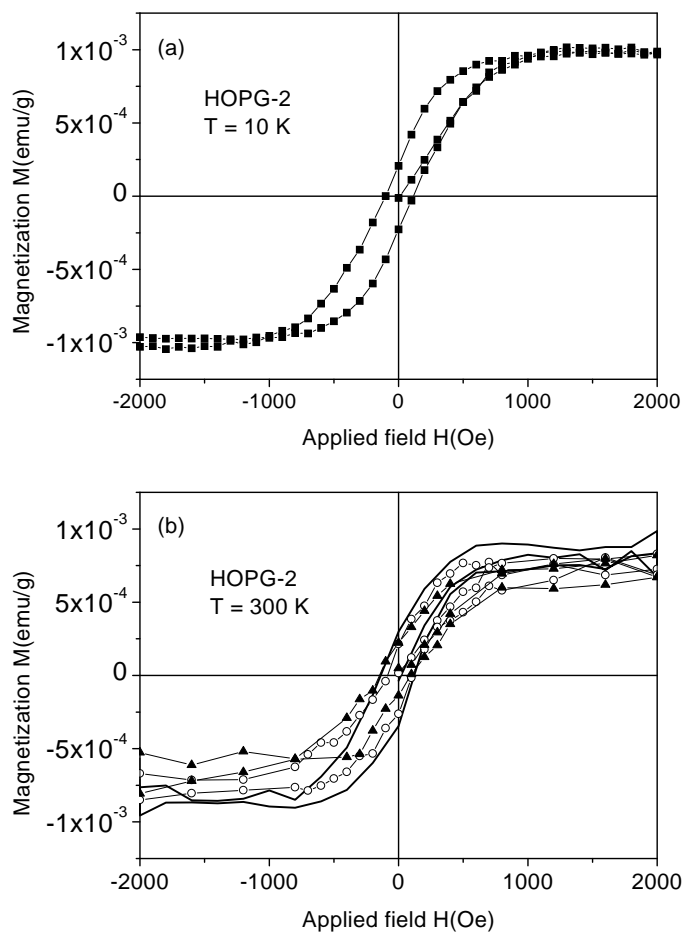}}} 
\caption{(a) Magnetization as a function of applied field at $T = 10~$K for sample HOPG-2. 
(b) Similar plot for a different piece of the same sample at 300~K after annealing in vacuum at a temperature of 700 K 
for 16 h $(\circ)$ and after 19 h
 at 800 K $(\blacktriangle)$. The bold  line represents the hysteresis at the same
temperature before annealing. A diamagnetic background has been subtracted.} 
\label{hopg2} 
\end{figure} 

\begin{figure} 
\centerline{\mbox{\epsfysize=12cm \epsffile{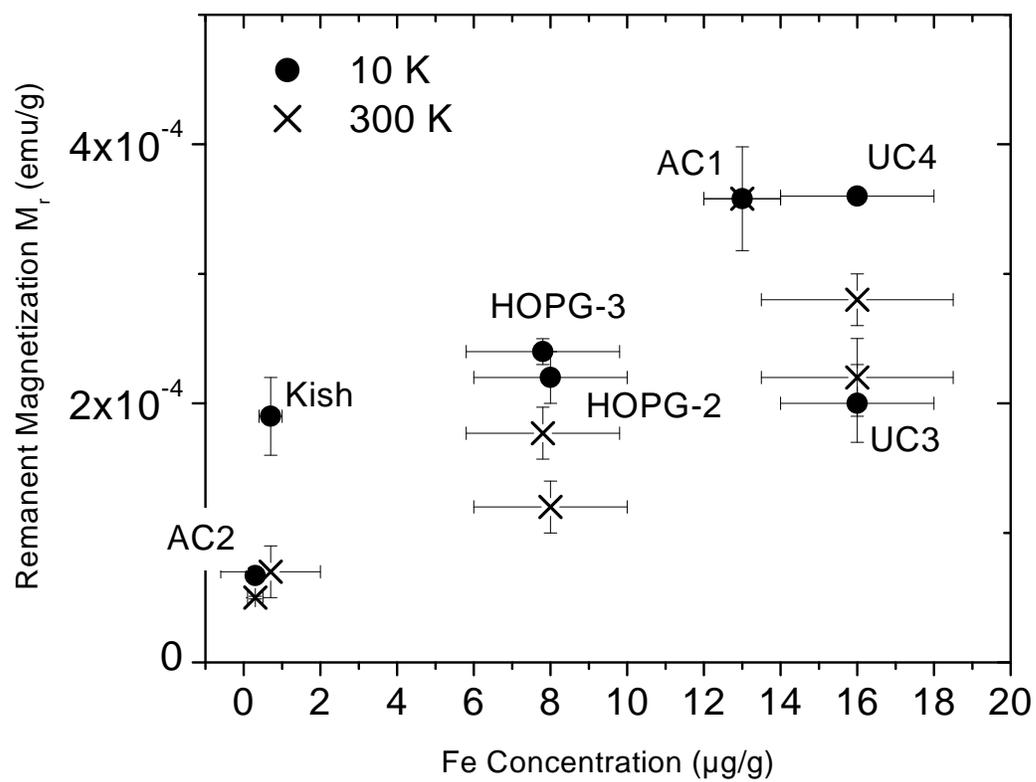}}} 
\caption{Remanent magnetization as a function of Fe concentration at two temperatures of
all HOPG and Kish graphite samples.} 
\label{rem} 
\end{figure}

\begin{figure} 
\centerline{\mbox{\epsfysize=12cm \epsffile{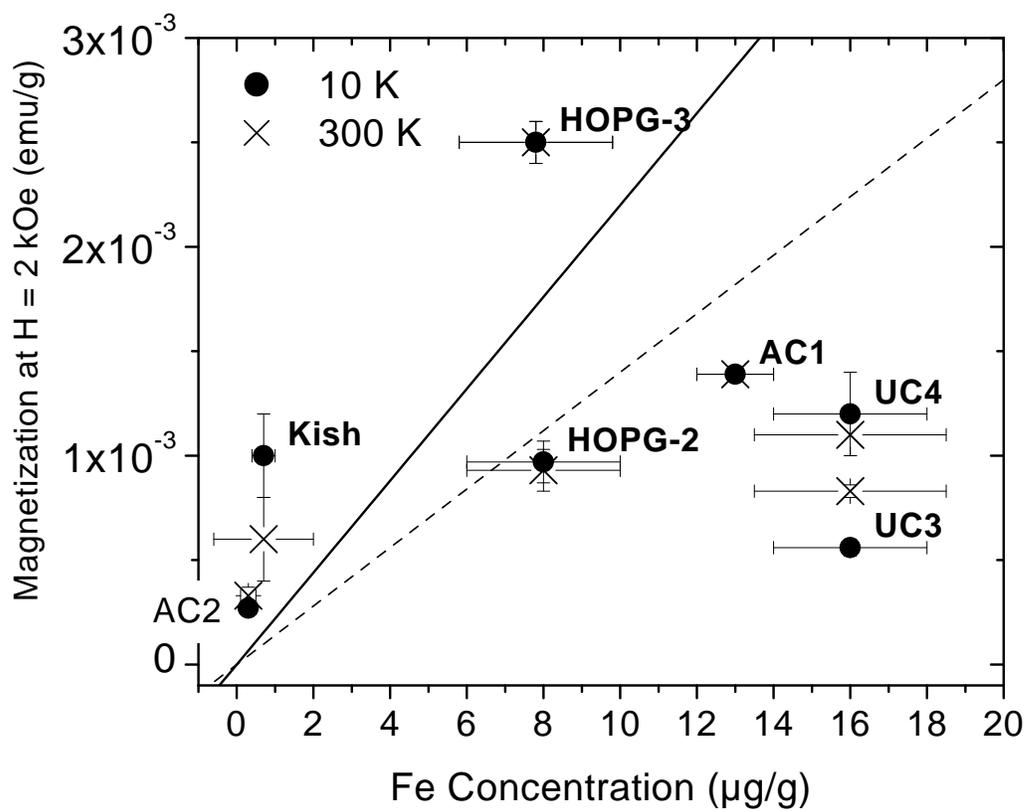}}} 
\caption{Magnetization at 2 kOe as a function of Fe concentration at two temperatures of
HOPG and Kish graphite samples. The solid line represents the expected magnetization 
if  Fe contained in the samples was in a ferromagnetic state. The dashed line represents a similar relation 
for Fe$_3$O$_4$.} 
\label{sat} 
\end{figure} 

\begin{figure} 
\centerline{\mbox{\epsfysize=12cm \epsffile{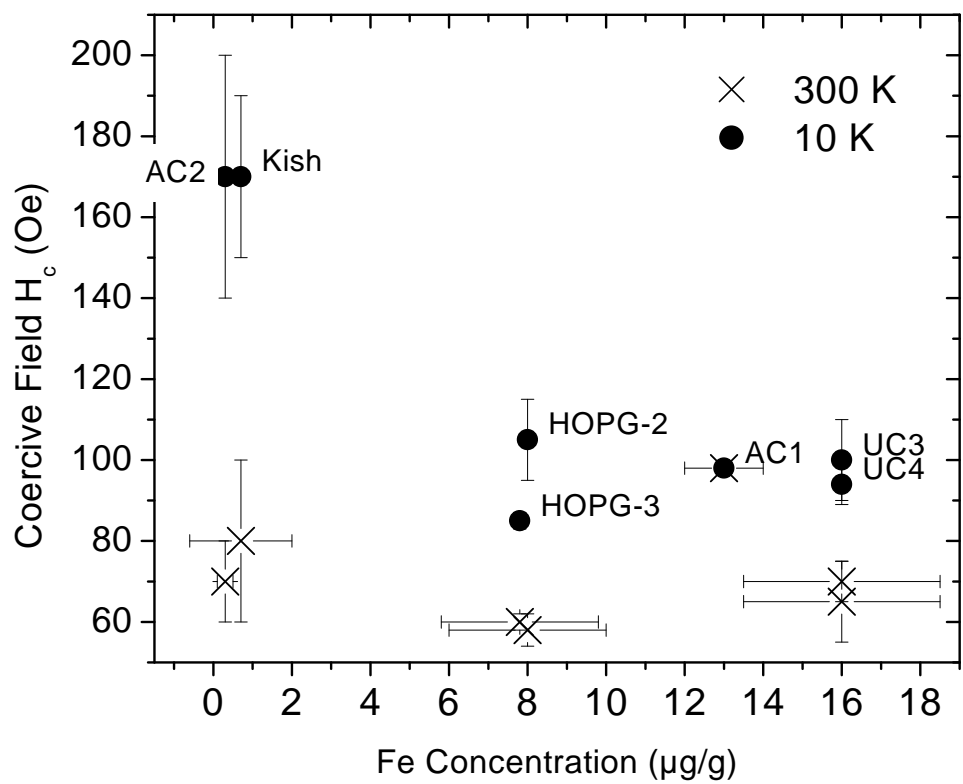}}} 
\caption{Coercive field at two temperatures of
the HOPG and Kish graphite samples as a function of Fe concentration.} 
\label{hc} 
\end{figure}

\begin{figure} 
\centerline{\mbox{\epsfysize=15cm \epsffile{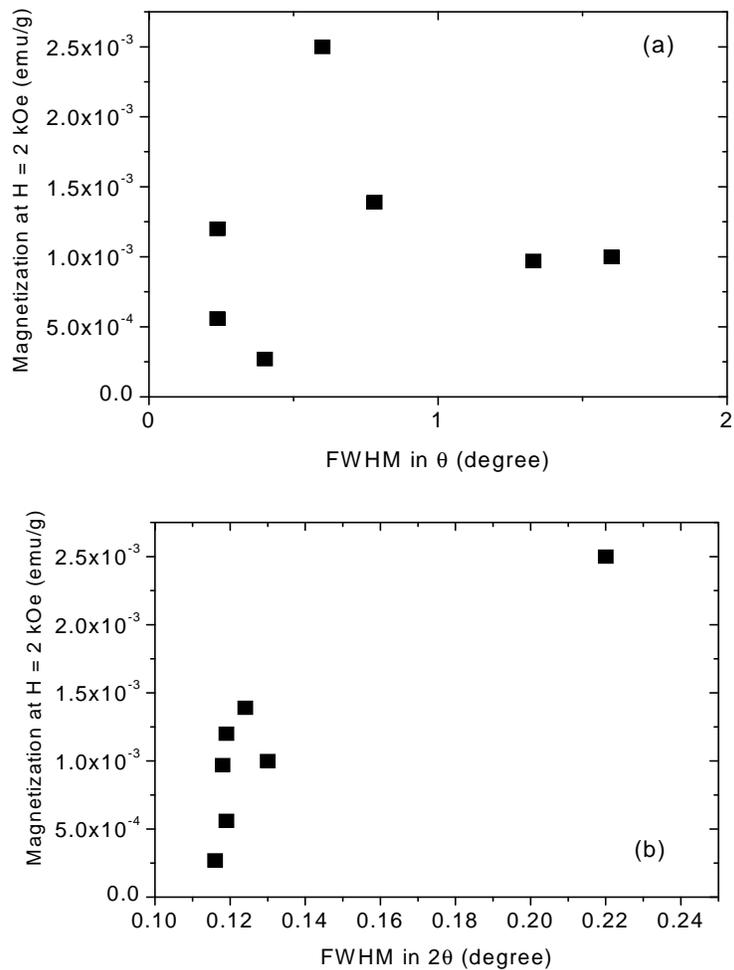}}} 
\caption{(a) Magnetization at 2kOe $(T = 10$~K) as a function of 
the full width at half maximum (FWHM) in $\theta$ for all HOPG 
and Kish graphite samples. (b) Similar plot as a function of the FWHM in $2\theta$.} 
\label{fwhm} 
\end{figure} 

\begin{figure} 
\centerline{\mbox{\epsfysize=15cm \epsffile{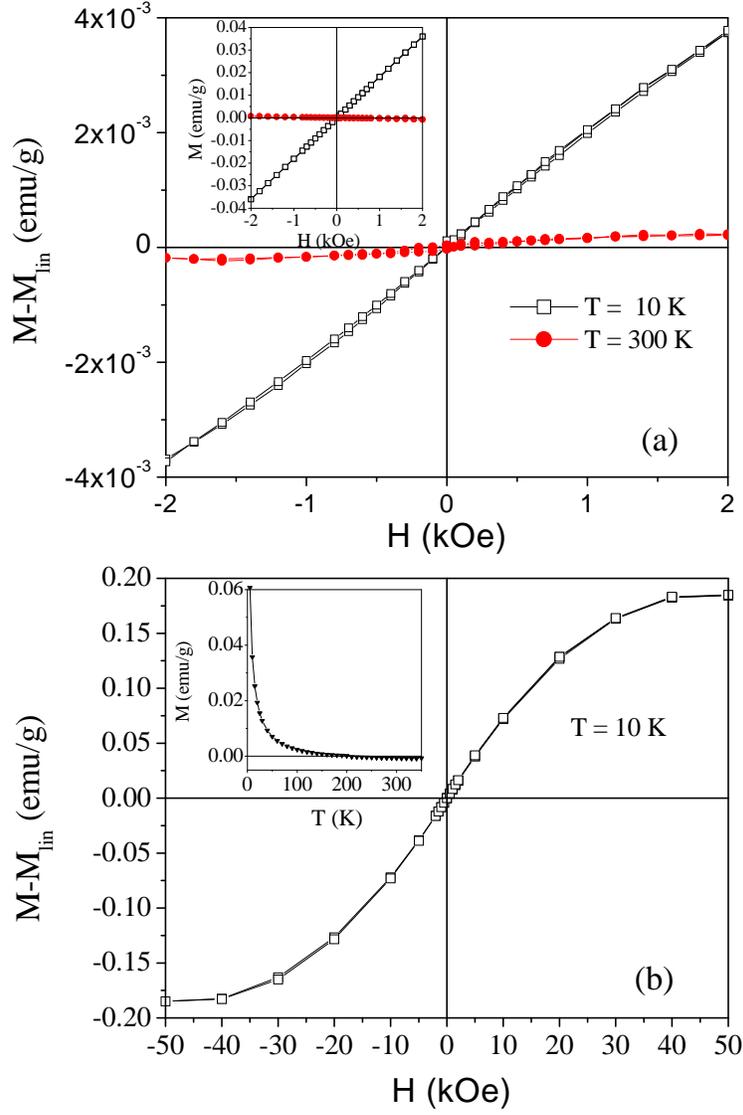}}} 
\caption{(a) Magnetization as a function of applied field at 10~K and 300~K for 
the natural graphite crystal with an average Fe-concentration of more than $500~\mu$g/g. 
A linear background contribution was subtracted. These
are shown in the inset. (b) Similar plot at 10 K and up to $\pm 50$~kOe applied field. The inset
shows the temperature dependence of the magnetization at 2 kOe.} 
\label{nat} 
\end{figure}

\begin{figure} 
\centerline{\mbox{\epsfysize=12cm \epsffile{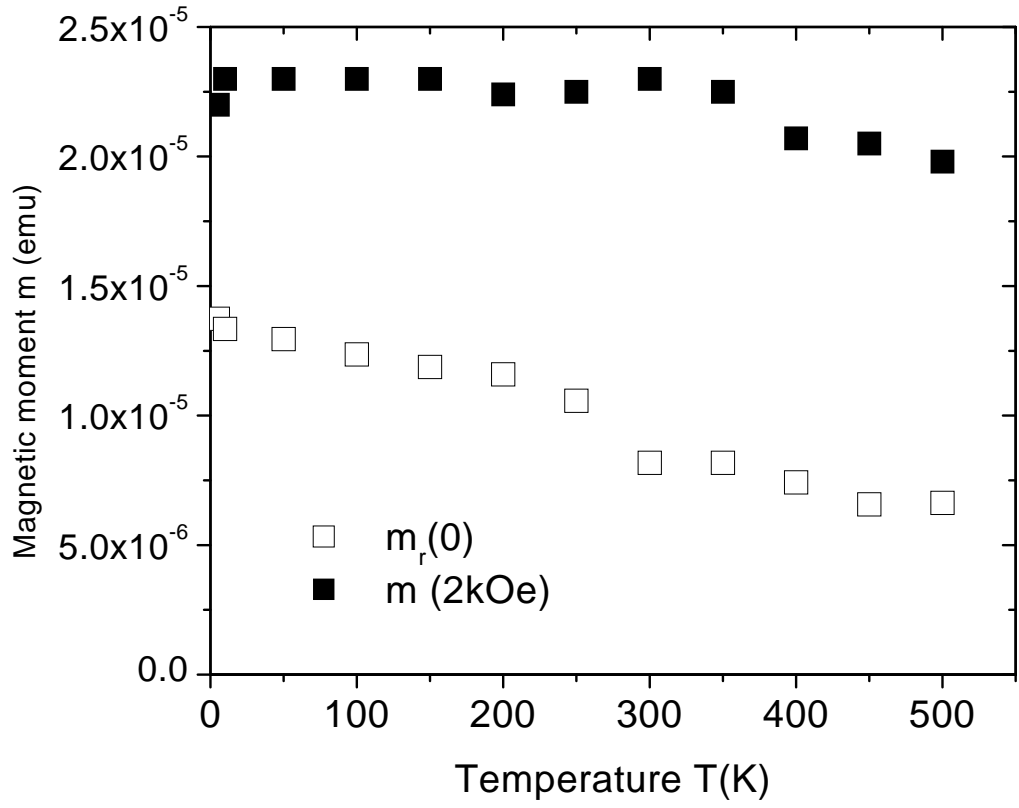}}} 
\caption{Temperature dependence of the saturation $(\blacksquare)$
and remanence $(\square)$ magnetic moment for the sample HOPG-2 after annealing 16 h at 700 K.
In the present work and due to technical reasons we
were not able to measure $M(T)$ continuously up to 800~K.} 
\label{magt} 
\end{figure}

\centerline{\mbox{\epsfxsize=14cm \epsffile{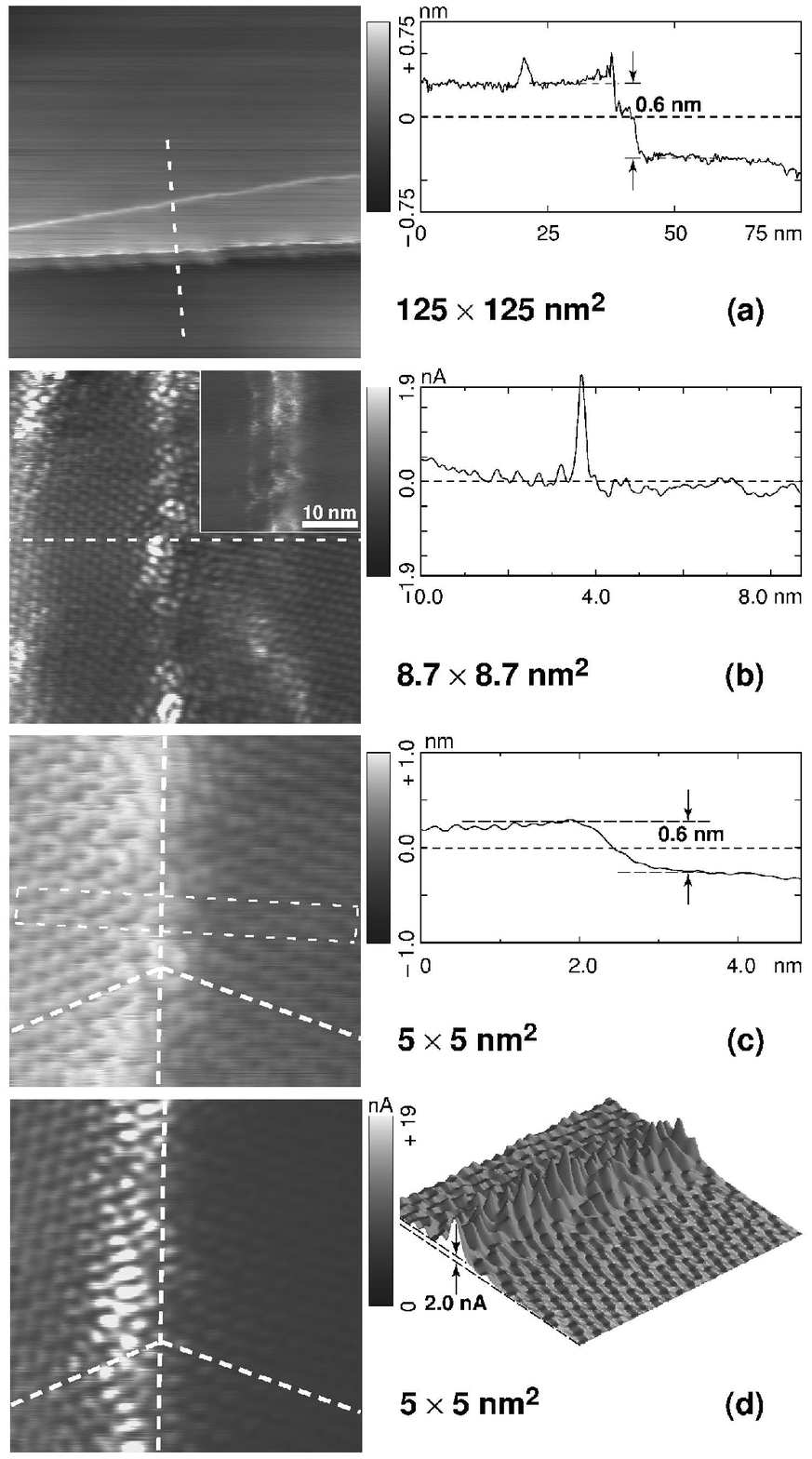}}} 

\begin{figure} 
\caption{STM images of a freshly cleaved surface of the sample, HOPG-2, in air 
(left) and their analysis using section profiles (right). The ordinate scales 
in (a) -- (c) apply to both the section profiles and the grey scales. 
(a) Overview scan in constant-current mode ($V_t = 32~$mV, $I_0 = 1.25~$nA, 
$f = 6~$Hz). In the section profile determined along the dashed line (right), 
a linear ramp has been subtracted to emphasize the flatness of the plateaus. 
(Thus, the profile levels do not entirely 
faithfully correspond to the grey levels shown in the micrograph.) 
Close to the step line that is recognized in the lower third of the image, an increase 
of the slope of the section profile is observed. Correcting for this overshoot, 
a step height of $\approx 0.6~$nm is determined. 
(b) High-resolution constant-height image and cross-section ($V_t = 33.5~$mV, 
$I_0 = 1.0~$nA, $f = 27.5~$Hz) of a surface region incorporating a line defect 
similar to the one shown in panel (a) ``north" of the step defect. 
Atomic resolution is achieved within the planar sections of the sample surface 
and indicates that the line defect corresponds to a boundary between differently oriented grains. The inset in the micrograph shows a 
constant-current ($V_t = 33.5~$mV, $I_0 = 1.0~$nA, $f = 2.0~$Hz) overview of 
the same surface feature. Scan size here is $25.4 \times 25.4~$nm$^2$. 
The grey scale of the overview is distinct from that of the main image and 
ranges from 0 to 1.0~nm (black to white).
(c) and (d) Constant-current  ($V_t = 20~$mV, $I_0 = 2.0~$nA, $f = 2.9~$Hz) and 
constant-height ($V_t = 20~$mV, $I_0 = 2.0~$nA, $f = 55~$Hz) images, respectively, 
of the step defect shown in panel (a). The scan direction has been rotated by 
$90^o$ with respect to that in the overview scan. The section profile in (c) 
has been determined in the rectangle bounded by thin dashed lines. Fat dashed 
lines indicate the grain boundary as well as the prevalent lattice directions 
within the grains near the boundary. In panel (d), the image is dominated by 
giant corrugations at atom sites immediately at the boundary. The relative 
proportions of these corrugations with respect to the amplitudes observed further 
away from the defect line is more clearly revealed in the pseudo-3D plot shown 
on the right.} 
\label{stm} 
\end{figure}

\end{widetext}

\end{document}